\documentclass[11pt,a4paper]{article}
\usepackage{jheppub}
\usepackage{physics}
\usepackage{color}
\title{Kink Crystalline Condensate and Multi-kink Solution in Holographic Superconductor}
                                           \author{Masataka Matsumoto, Shin Nakamura, and Ryosuke Yoshii}
                                           \affiliation{Department of Physics, Chuo University,\\
                                           Bunkyo-ku, Tokyo 112-8551, Japan}
                                           \emailAdd{matumoto@phys.chuo-u.ac.jp, nakamura@phys.chuo-u.ac.jp, yoshii@phys.chuo-u.ac.jp}
                                           \abstract{
The theory of superconductivity can be divided into two groups depending on whether it has multi-kink solutions.
For example, the BCS theory and the Gross-Neveu model have metastable multi-kink solutions whereas the conventional Ginzburg-Landau theory without higher-derivative interactions does not have any multi-kink solutions.
In this paper, we systematically examine the solutions of the holographic superconductor model to find out which group the model falls into.
We show that the holographic superconductor model has metastable multi-kink solutions.
In this sense, we find that the holographic superconductor model falls into the category of the BCS theory and the Gross-Neveu model.
We also find that the holographic superconductor model has kink crystalline condensates which are well-fitted by the Jacobi elliptic functions.
}                                        
                                           \keywords{AdS-CFT Correspondence, Gauge-gravity correspondence, Holography and condensed matter physics(AdS/CMT), Solitons Monopoles and Instantons}
                                           \arxivnumber{}
                                           \begin{document}
                                           \maketitle
                                           \flushbottom
\section{Introduction}           
Recently, strongly coupled systems have been investigated thanks to the Gauge/Gravity correspondence\,\cite{Maldacena1997,Gubser1998,Witten1998}. In particular, this correspondence has been applied to wide area of the condensed matter physics\,(for example, see Refs.\,\cite{Ammon2015,Natsuume2014,Zaanen2015,Hartnoll2016}). One of the important applications is the holographic superconductor\,\cite{Hartnoll2008,Hartnoll2008k}. The spontaneous symmetry breaking of the $U(1)$ symmetry has been presented in Ref.\,\cite{Gubser2008} for an Abelian Higgs model coupled to the gravity theory with a negative cosmological constant, and the dual description of the superconductivity has been further explored in\,\cite{Hartnoll2008,Hartnoll2008k}. The holographic superconductor was also studied in the presence of an external magnetic field\,\cite{Nakano2008,Albash2008} or a current\,\cite{Basu2008,Herzog2008,Arean2010}.
Another interesting topic in the holographic superconductor is the spatially inhomogeneous structure of the condensate and the charge density\,\cite{Keranen2009,Keranen22009,Keranen32009,Lan2017,Flauger2010,Cremonini2016}. These studies are of particular interest to the inhomogeneous structure in strongly correlated electron systems. Moreover, the analysis on the inhomogeneous solutions might yield the deeper insight on the dual field theory of the holographic superconductor.
However, the possible solutions in holographic superconductor have not yet been explored systematically.

In general, inhomogeneous solutions are found in various models. However, these solutions are realized under the specific conditions.
In condensed matter physics, it has been proposed that strongly correlated superconductors exhibit various inhomogeneous phases, such as charge or spin density wave phase, and the exotic state which is called Larkin-Ovchinnikov-Fulde-Ferrel\,(LOFF) phase\,\cite{LO,FF}. For example, it is expected that the LOFF phase is stabilized at low temperatures and in high magnetic fields\,\cite{Klein2004,Zyuzin2009,Aoyama2013,Quan2010,Yoshii2015}. The indirect experimental evidences of the LOFF phase have recently been reported in a heavy fermion compound\,\cite{Radovan2003}, an organic superconductor\,\cite{Yonezawa2008}, or ultra-cold atomic gases\,\cite{Liao2010}. In addition, topological solitons in conducting polymer\,\cite{Brazovskii1,Brazovskii2,Brazovskii3,Campbell1982,Heeger1988} and interesting phenomena such as the fermion number fractionization\,\cite{Jackiw1975,Niemi1984} associated with that have been found.
In QCD, the spatially modulated chiral condensate has been studied in order to understand the rich phase structure of QCD\,(see Refs.\,\cite{Casalbuoni2003,Buballa2014} for reviews). This feature has been investigated in the (1+1) dimensional Nambu--Jona-Lasinio model or the chiral Gross-Neveu model\,\cite{Nambu1961,Gross1974,Basar2008,Basar2008k,Basar2009}. Recently, the connection between the Nambu--Jona-Lasinio model and the non-linear sigma model is also of particular interest to the inhomogeneous condensate\,\cite{Yoshii2018}. Moreover, the generalized Ginzburg-Landau approach with higher derivatives of the order parameter has been developed and it has been revealed that the LOFF phase of the chiral condensate is energetically preferred in Ref.\,\cite{Nickel2009}.

Since the proposal of the holographic superconductor model, various aspects of superconductivity are rederived. However, it is still unclear how far it captures the physics of superconductivity. So far, though this model is known to be consistent with the Ginzburg-Landau theory, it is not clear whether it can describe the physics beyond the Ginzburg-Landau theory or not. 

The theory of superconductivity can be divided into two groups depending on whether it has multi-kink solutions.
For example, the microscopic models such as the BCS theory and the Gross-Neveu model have multi-kink solutions. They are metastable and are thermodynamically unfavored compared to the homogeneous solutions \cite{Yoshii2015}.
On the other hand, the conventional Ginzburg-Landau theory without higher-derivative interactions does not have the multi-kink solutions \cite{Takahashi2012}.

In this paper, we systematically examine the solutions of the holographic superconductor model to find out which group the model falls into.
Our work is an extension of the previous reports on holographic dark solitons\,\cite{Keranen2009,Keranen22009,Lan2017} and we focus on inhomogeneous solutions with constant chemical potential in space. 
We find that the holographic superconductor model has multi-kink solutions. Computing the thermodynamic potential of these solutions, we find that they are metastable and are thermodynamically unfavored compared to the homogeneous solutions.
In this sense, we conclude that the holographic superconductor model falls into the category of the BCS theory and the Gross-Neveu model.
This suggests that the holographic superconductor model captures phenomena beyond the conventional Ginzburg-Landau model.

Furthermore, we find that the holographic superconductor model has the kink crystalline condensates of the LO-like phase which are well-fitted by the Jacobi elliptic functions.

This paper is organized as follows. In Sec.\,\ref{sec:2}, we briefly review the holographic superconductor. We introduce the holographic superconductor model and review the derivation of homogeneous solutions and its extension to the inhomogeneous case. We obtain the solution with kink crystalline condensate and the multi-kink solution. We find that the charge density becomes minimum at the location of condensate's node. In Sec.\,\ref{sec:3}, we compute the thermodynamic potential of solutions. We find that inhomogeneous solutions have lower value of the thermodynamic potential than that of the normal state and higher values than that of the homogeneous solution. 
In Sec.\,\ref{sec:5}, we present conclusion and discussion.

\section{Holographic Superconductor}
\label{sec:2}
\subsection{Holographic Setup}
To begin with, we review the basic setup of the holographic superconductor. We consider the Einstein-Maxwell theory with a negative cosmological constant and a charged complex scalar field $\Psi$ in (3+1) dimensional spacetime\,\cite{Gubser2008,Hartnoll2008,Hartnoll2008k}. The action of this model is given by
\begin{equation}
	S=\int d^4x \sqrt{-g}\left[ \frac{1}{16 \pi G_{N}} \left(R+6 \right)-\frac{1}{4}F_{\mu\nu}F^{\mu\nu}-\left| D_{\mu} \Psi \right|^{2} +V\left( \left| \Psi \right| \right)  \right].
\end{equation}
The covariant derivative is defined by $D_{\mu}\Psi=\left( \nabla_{\mu} -iqA_{\mu} \right)\Psi$, where $A_{\mu}$ is the $U(1)$ gauge field and $q$ is the charge of the scalar field. Here, the AdS radius is set to be 1. $\mu$ and $\nu$ are indices of the four dimensional bulk spacetime. This bulk system with a $U(1)$ gauge symmetry is dual to the strongly coupled field theory with a global $U(1)$ symmetry in (2+1) dimensions\,\cite{Horowitz2010}. If one rescales the scalar field and the gauge field by $q$ and takes the limit $q \rightarrow \infty$ with the amplitude of fields fixed, the matter sector is decoupled from the gravity sector. In this limit, the so-called probe limit, one can ignore the backreaction of the matter sector to the background geometry. Then the Einstein equations yield the (3+1) dimensional planar AdS black hole geometry
\begin{equation}
    ds^{2}=\frac{L^{2}}{z^{2}} \left[ -f(z) dt^{2} + \frac{dz^{2}}{f(z)} + dx^{2}+dy^{2} \right],
    \label{eq:metric}
\end{equation}
where
\begin{equation}
    f(z)=1-\left( \frac{z}{z_{H}} \right)^{3}.
\end{equation}
Here, $z_{H}$ is the location of the black hole horizon. The Hawking temperature is given by $T_{H}=3/(4 \pi z_{H})$, which corresponds to the temperature of the dual field theory. 
The dynamics of this system is determined by the action of the matter sector
\begin{equation}
	S_{\rm matter}=\int d^4x \sqrt{-g}\left[ -\frac{1}{4}F_{\mu\nu}F^{\mu\nu}-\left| D_{\mu} \Psi \right|^{2} +V\left( \left| \Psi \right| \right)  \right].
	\label{eq:action}
\end{equation}
We assume  that the potential is given by $V=-m^{2} \left|\Psi \right|^{2}$. For simplicity, we set $q=z_{H}=1$. The equation of motion for the complex scalar field is given by
\begin{eqnarray}
    0&=&\frac{1}{\sqrt{-g}} D_{\mu} \left( \sqrt{-g} D^{\mu}\Psi\right) -m^{2} \Psi.
\end{eqnarray}
The Maxwell equations are obtained as
\begin{eqnarray}
    \frac{1}{\sqrt{-g}} \partial_{\mu} \left( \sqrt{-g} F^{\mu\nu}\right) =i (\Psi^{*} \partial^{\nu} \Psi-\Psi \partial^{\nu} \Psi^{*}) - 2A^{\nu}\left| \Psi \right|^{2}.
    \label{Maxwelleq}
\end{eqnarray}

\subsection{Homogeneous Solutions}
In this subsection, we review the homogeneous case of the holographic solution by following Ref.\,\cite{Hartnoll2008}. First, we take the ansatz that the scalar field and the gauge field, respectively, have the following form
\begin{equation}
    \Psi=\Psi(z), \;\;\; A=A_{t}(z) dt,
\end{equation}
where we employed the gauge $A_{x}=A_{y}=A_{z}=0$. Since the phase of $\Psi$ must be constant in space in this gauge from the Maxwell equations (\ref{Maxwelleq}), we take $\Psi$ to be real. Then, the equation of motion for the scalar field becomes
\begin{equation}
    \Psi''+\left(\frac{f'}{f} -\frac{2}{z} \right)\Psi' + \left( \frac{A_{t}^{2}}{f^{2}}-\frac{m^2}{f z^{2}}
 \right)\Psi=0,
 \label{eq:Psieqhomo}
\end{equation}
and the Maxwell equation becomes
\begin{equation}
    A_{t}''-\frac{2\Psi^{2}}{f z^{2}}A_{t}=0.
    \label{eq:Ateqhomo}
\end{equation}
Hereafter, we use the notation $y'=\frac{dy}{dz}$ and $y''=\frac{d^{2}y}{dz^{2}}$. We choose the mass of the scalar field to be $m^{2}=-2$, which is above the Breitenlohner-Freedman\,(BF) bound $m_{\rm BF}^{2}=-9/4$ for stability of AdS spacetime\,\cite{BF}. 


Let us consider boundary conditions in order to solve these equations of motion. The asymptotic forms of each field are obtained from these equations of motion at the vicinity of the AdS boundary,
\begin{equation}
    \Psi(z)=\Psi^{(1)} z + \Psi^{(2)} z^{2} +\cdots, \;\;\; A_{t}(z) = \mu -\rho z + \cdots.
    \label{eq:asymptotic}
\end{equation}
According to the AdS/CFT dictionary, $\mu$ and $\rho$ correspond to the chemical potential and the charge density, respectively. The condensate is defined by $\expval{{\cal O}_{2}}=\sqrt{2}\Psi^{(2)}$. For the scalar field, both of these two leading terms are normalizable since we choose a scalar mass close to the BF bound. There are two different possibilities to identify the source and the condensate of operators in the dual field theory. Since we are interested in the non-zero vacuum expectation value for the condensate without the corresponding source, there are two possible boundary conditions, $\Psi^{(1)}=0$ or $\Psi^{(2)}=0$. In this study, we focus on the condition $\Psi^{(1)}=0$. Thus, we obtain boundary conditions at the AdS boundary:
\begin{equation}   
    \Psi^{(1)}=0, \;\;\; A_{t}(0)=\mu.
    \label{eq:bdrycond1}
\end{equation}
 In addition, we impose regularity conditions at the black hole horizon as boundary conditions:
\begin{equation}
    \Psi'(1)=\frac{2}{3}\Psi(1), \;\;\; A_{t}(1)=0.
    \label{eq:bdrycond2}
\end{equation}
One can obtain these conditions by expanding Eqs.\,(\ref{eq:Psieqhomo}) and (\ref{eq:Ateqhomo}) with respect to $z$ near black hole horizon and assuming the regularity at the black hole horizon\,($z=1$). 

\begin{figure}[tbp]
\includegraphics[width=7.5cm,bb=0 0 360 264]{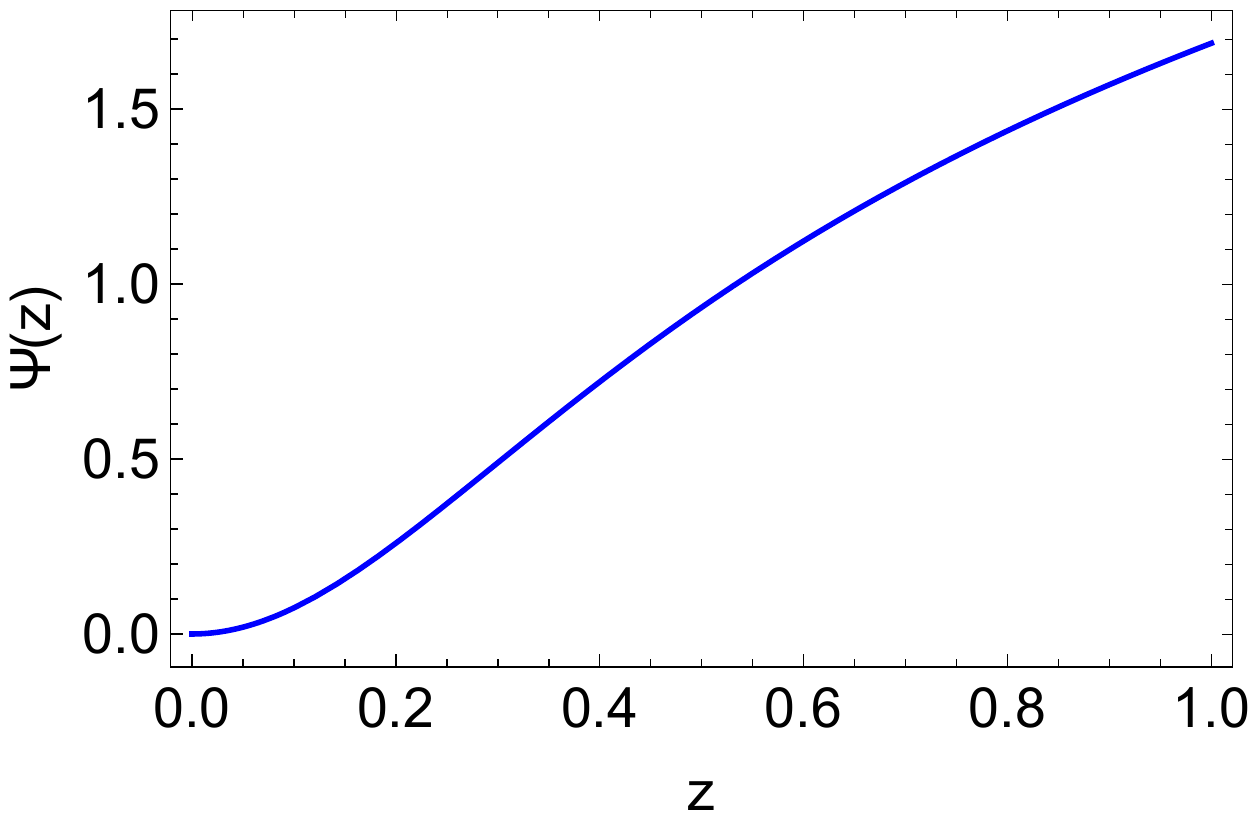}
\includegraphics[width=7.5cm,bb=0 0 360 264]{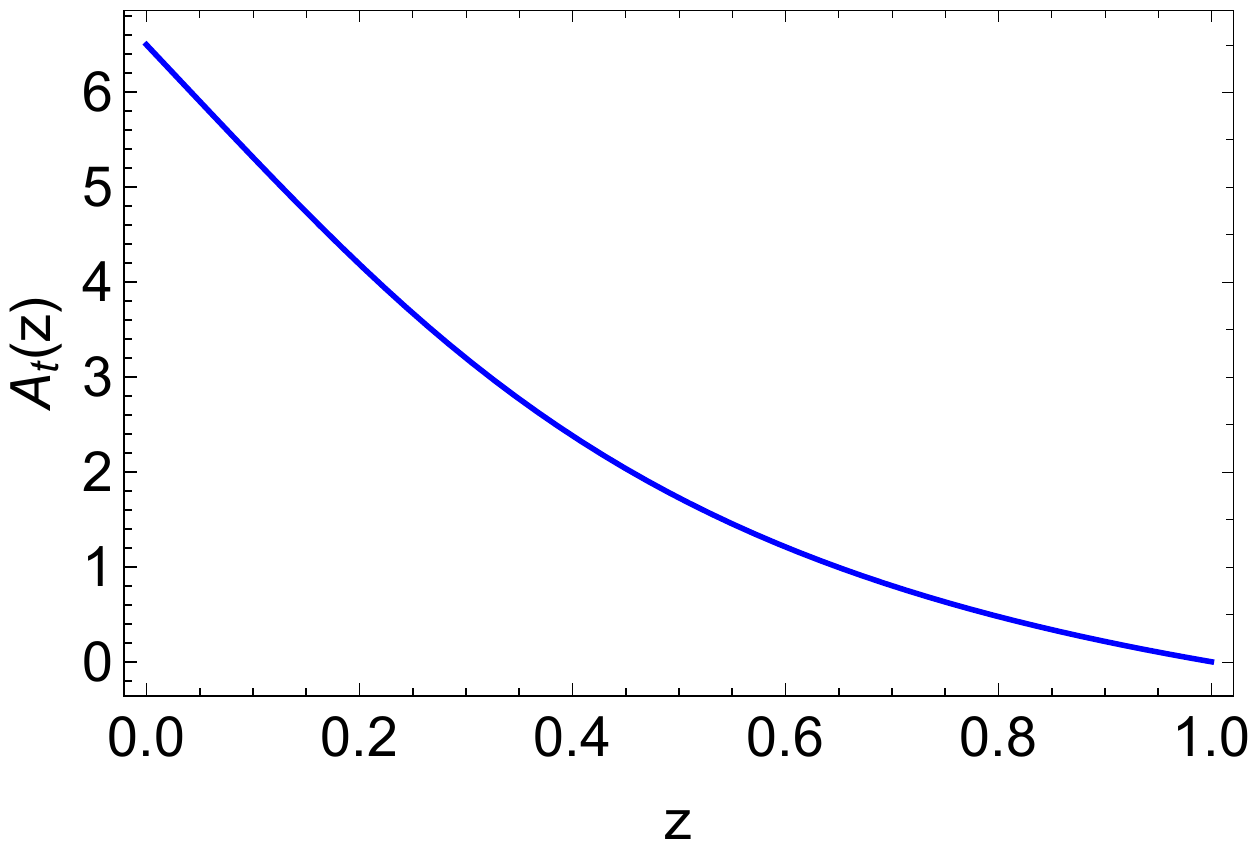}
\caption{$\Psi(z)$ and $A_{t}(z)$ for $\mu=6.5$.}
\label{fig:homo}
\end{figure}

Note that one of the solutions which satisfy these boundary conditions is $\Psi=0$ and $A_{t}=\mu(1-z)$. In this trivial solution, there is no scalar hair and thus this solution is referred to as the {\it normal state}. On the other hand, a non-trivial solution $\Psi\neq0$ is referred to as the {\it superconducting state}. In order to find a non-trivial solution, we have to solve Eqs.\,(\ref{eq:Psieqhomo}) and (\ref{eq:Ateqhomo}) numerically with above the boundary conditions (\ref{eq:bdrycond1}) and (\ref{eq:bdrycond2}). As an example of non-trivial solutions, we plot the scalar field $\Psi(z)$ and the gauge field $A_{t}(z)$ with $\mu=6.5$ in Fig.\,\ref{fig:homo}. The condensate $\expval{{\cal O}_{2}}$ and the charge density $\rho$ can be computed from asymptotic form of these solutions. A finite condensate implies that the black hole in the bulk has developed scalar hair.

\begin{figure}[tbp]
\includegraphics[width=7.4cm, bb=0 18 360 291]{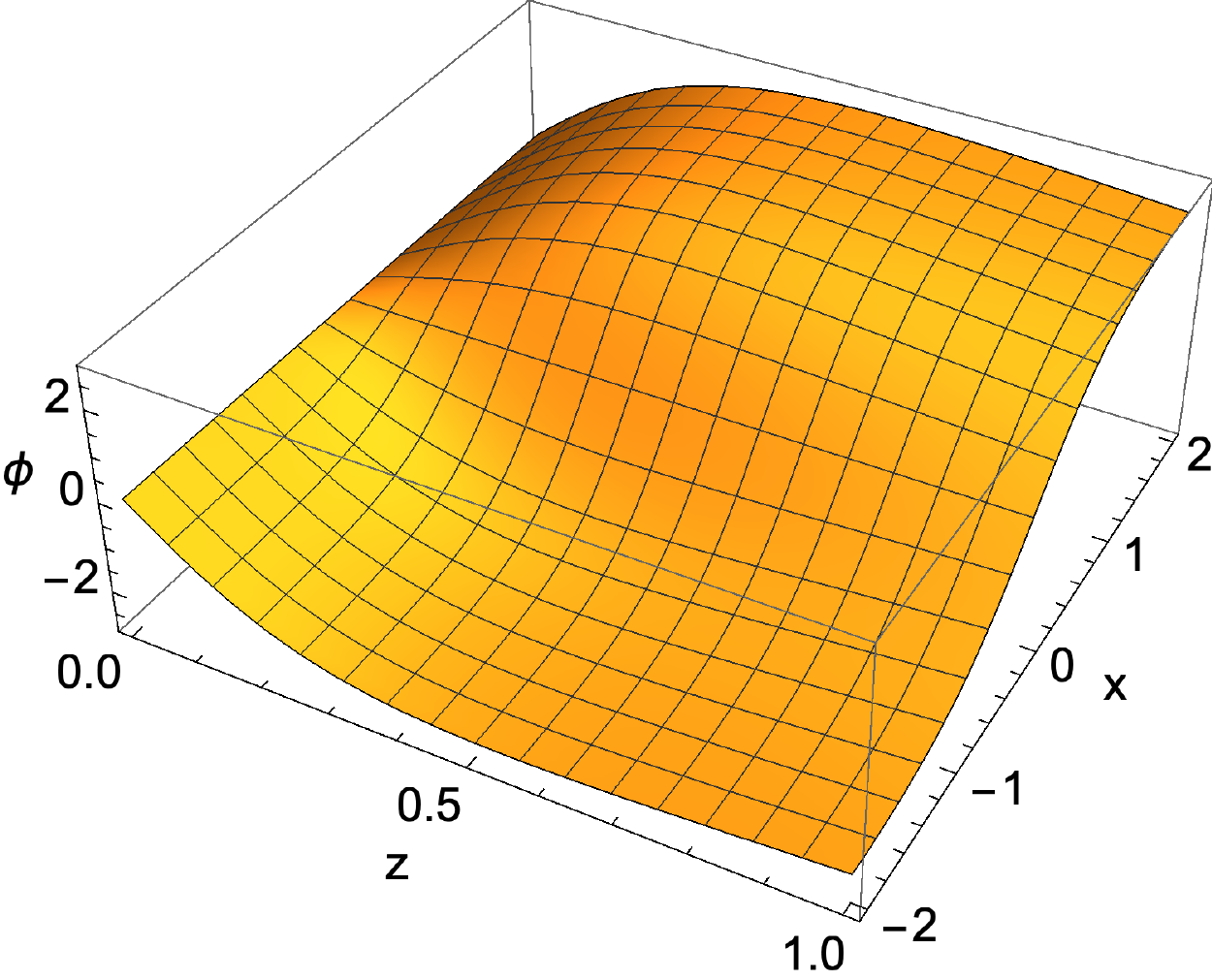}
\includegraphics[width=7.4cm, bb=0 18 360 291]{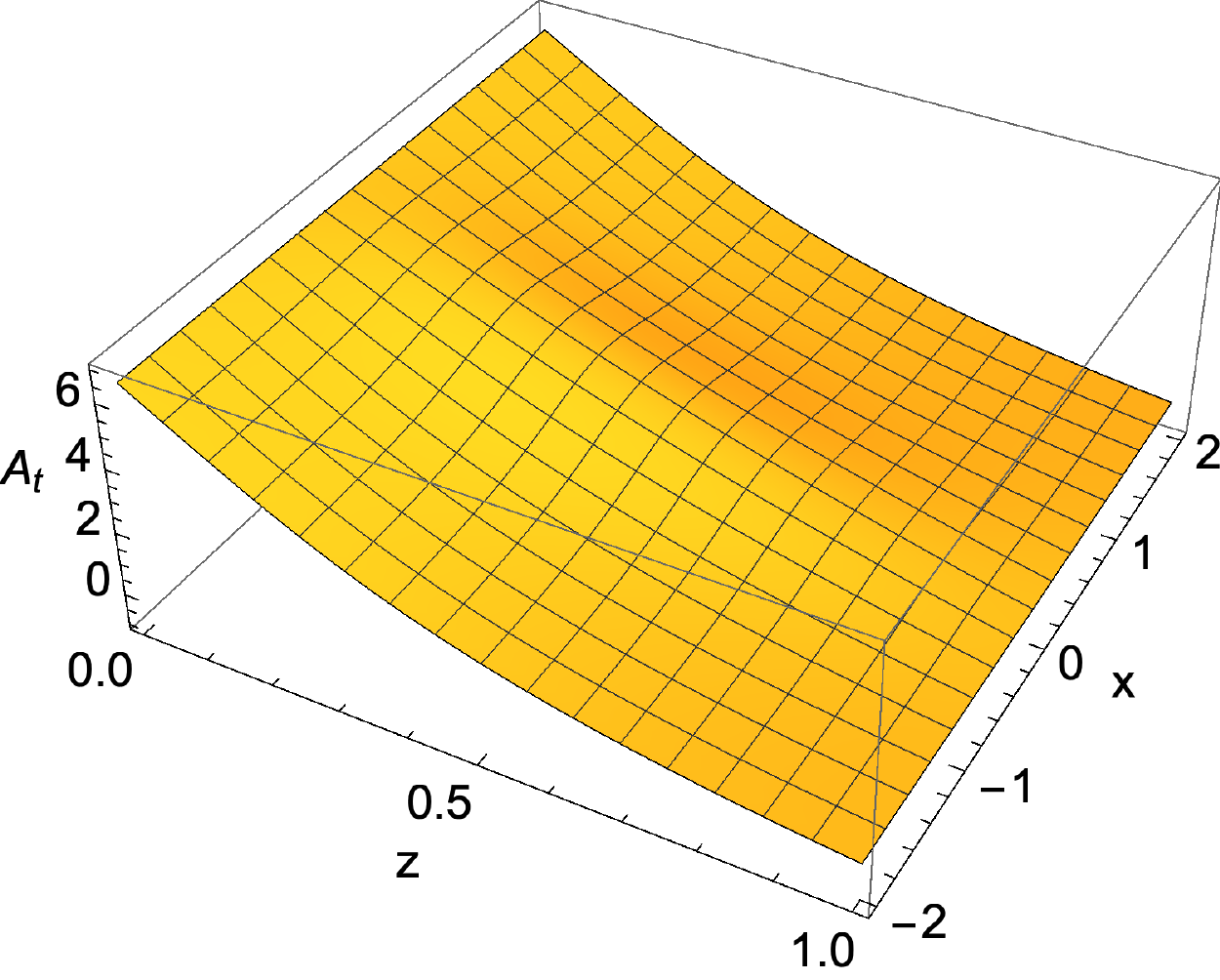}
\caption{Inhomogeneous configurations of the bulk solutions of $\phi(z,x)$ (left) and $A_{t}(z,x)$ (right) with single node on the $x$-axis for $\mu=6.5$.}
\label{fig:3D}
\end{figure}

\begin{figure}[tbp]
\includegraphics[width=7.4cm, bb=0 18 360 291]{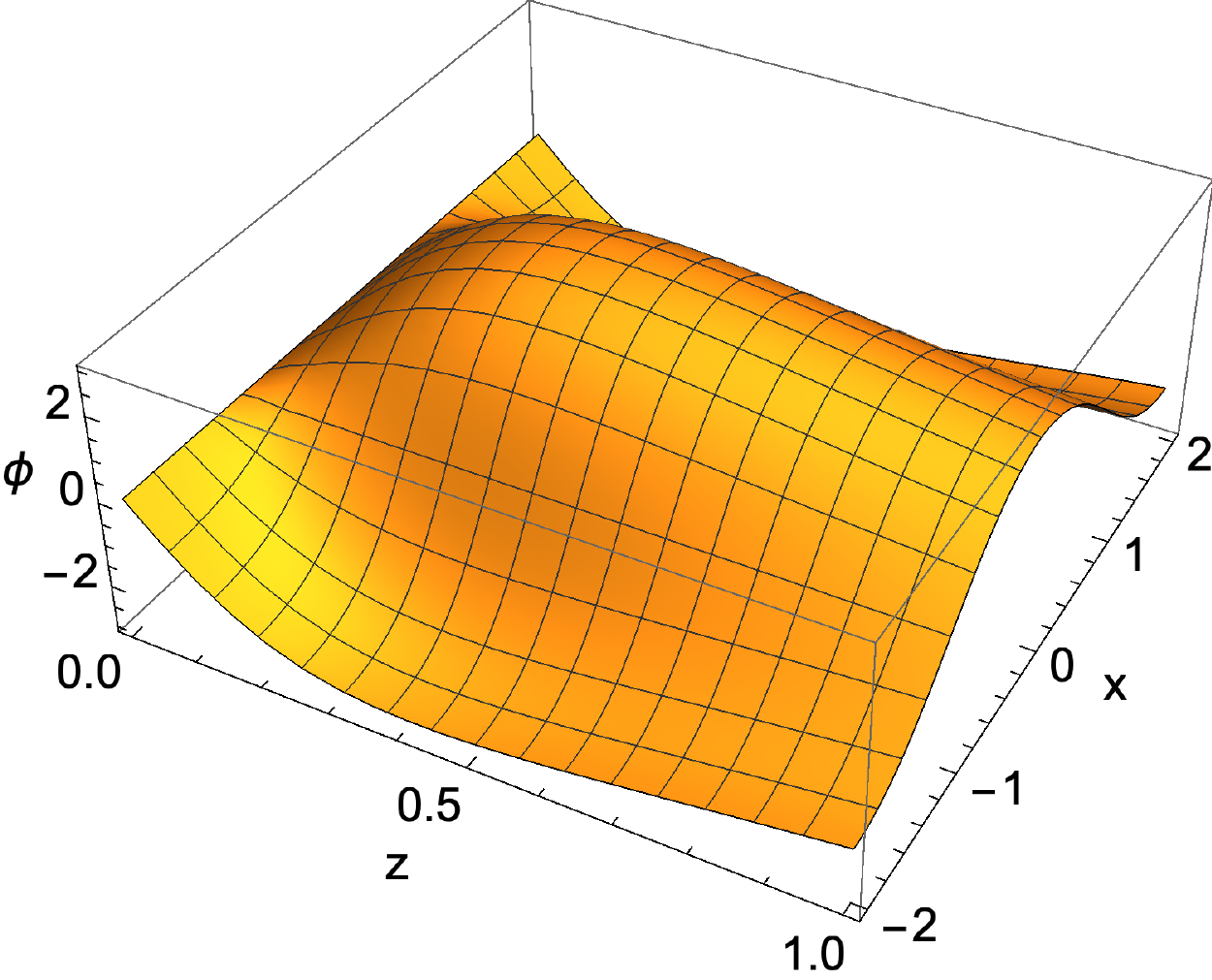}
\includegraphics[width=7.4cm, bb=0 18 360 291]{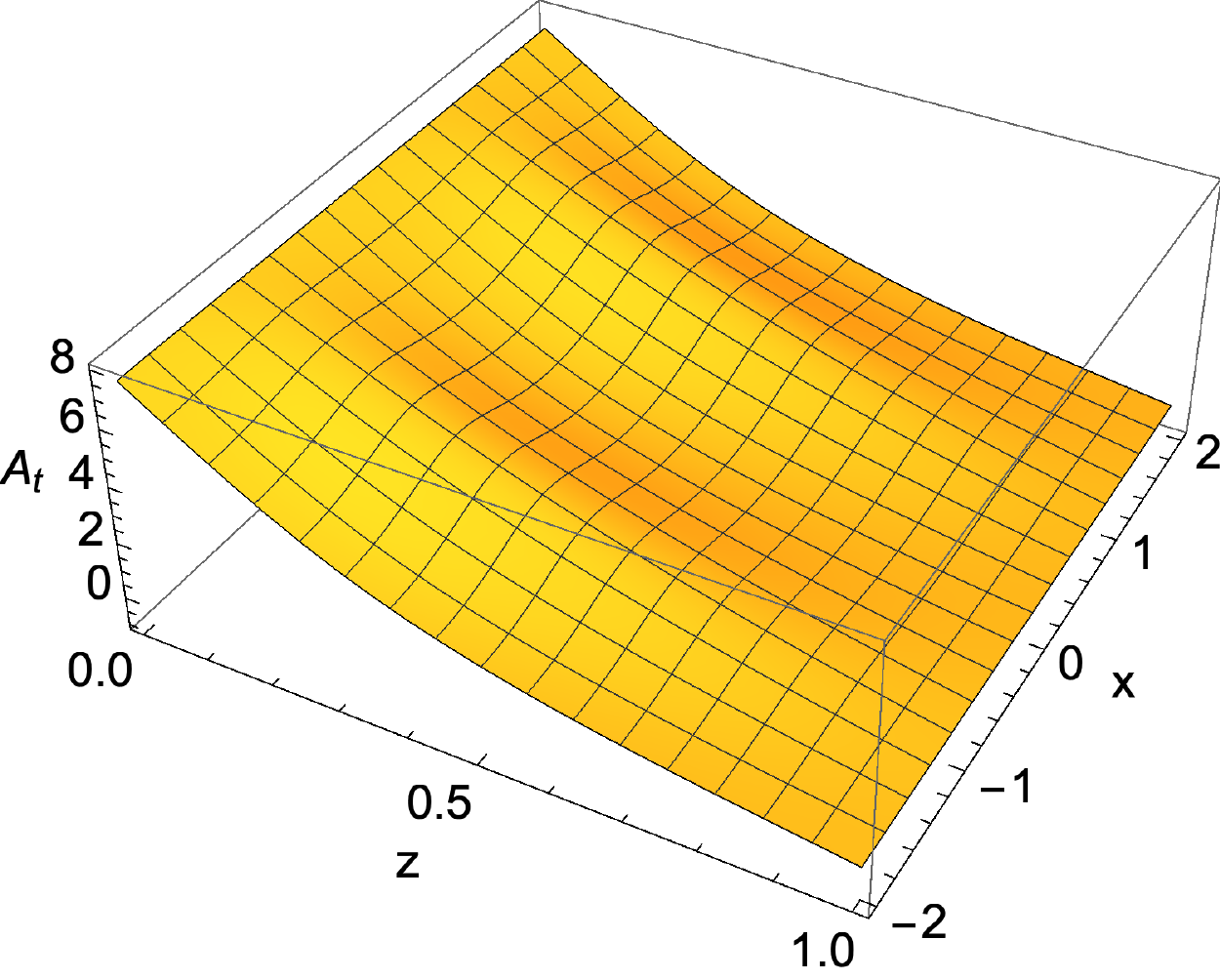}
\caption{Inhomogeneous configurations of the bulk solutions of $\phi(z,x)$ (left) and $A_{t}(z,x)$ (right) with two nodes on the $x$-axis for $\mu=7.5$.}
\label{fig:3D2}
\end{figure}

\begin{figure}[tbp]
\includegraphics[width=7.4cm, bb=0 18 360 291]{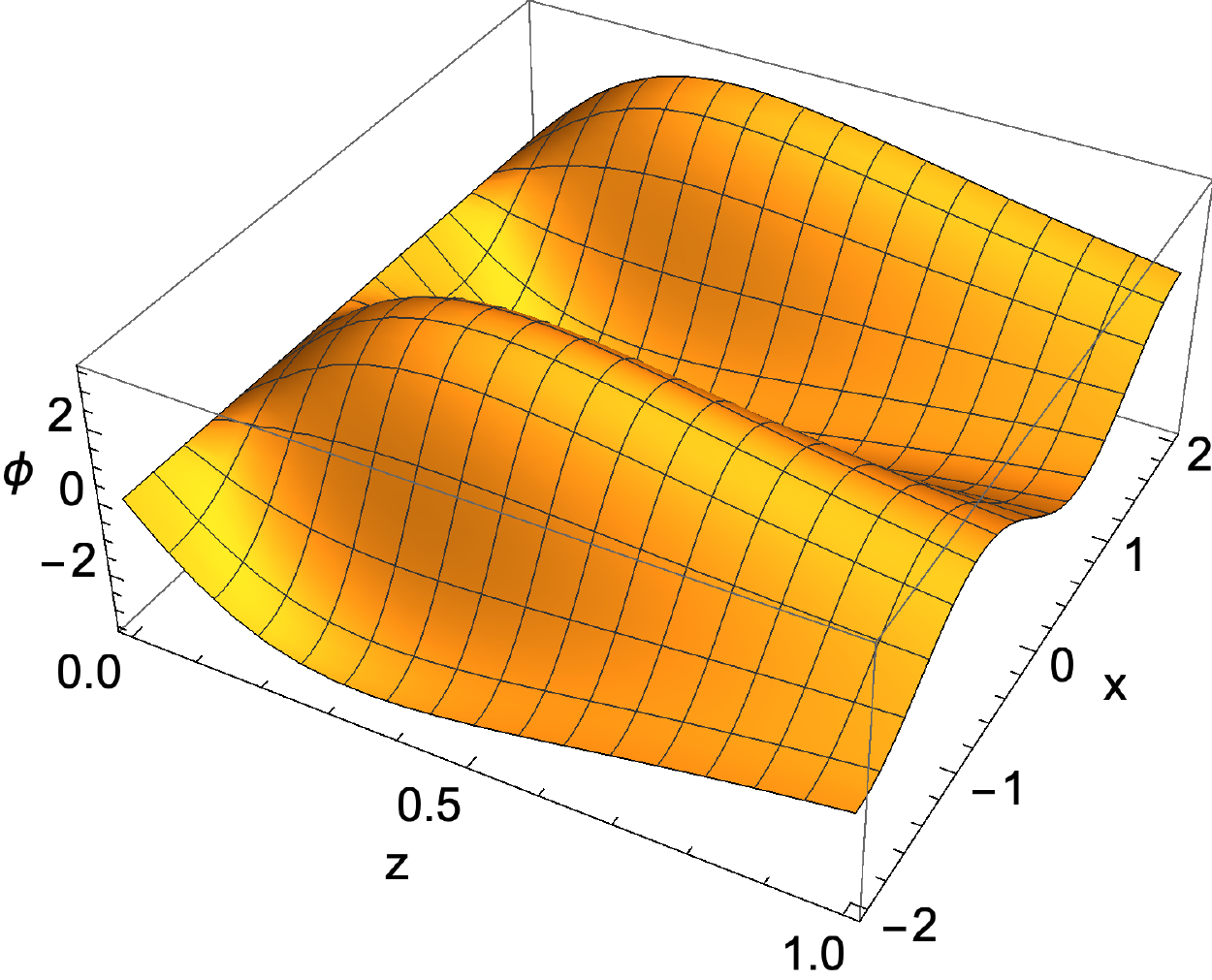}
\includegraphics[width=7.4cm, bb=0 18 360 291]{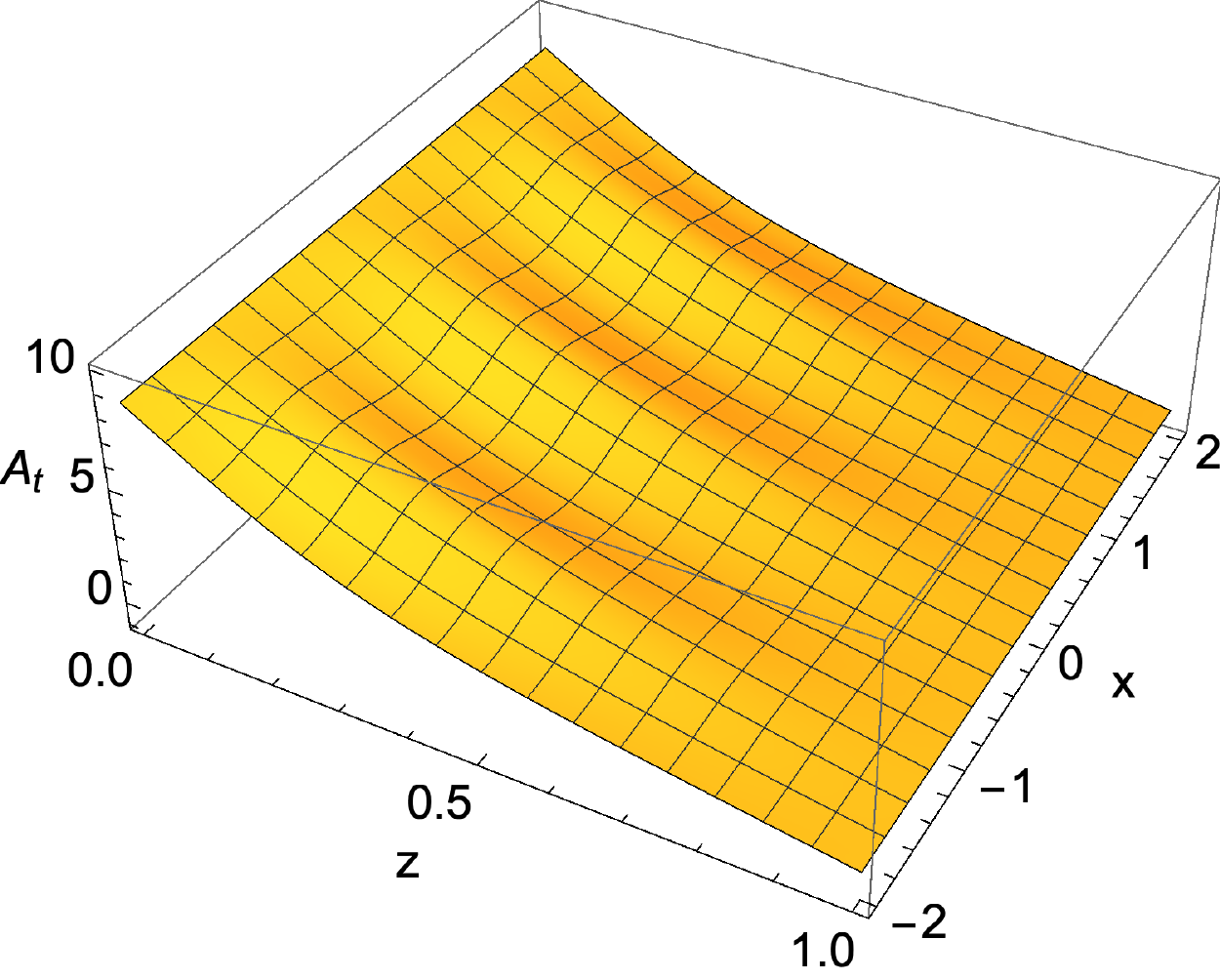}
\caption{Inhomogeneous configurations of the bulk solutions of $\phi(z,x)$ (left) and $A_{t}(z,x)$ (right) with three nodes on the $x$-axis for $\mu=8.5$.}
\label{fig:3D3}
\end{figure}

\subsection{Inhomogeneous Solutions}
In this subsection, since we are interested in the inhomogeneous solutions, we assume that the fields depend on the coordinates $z$ and $x$. The only non-zero component of the gauge field is $A_{t}=A_{t}(z,x)$. In the same gauge choice as the homogeneous case, one can set $\Psi$ to be real. For convenience, we define the field $\Psi(z,x)=z \phi(z,x) /\sqrt{2}$, where $\phi(z,x)$ is the real function. Thus, equations of motion can be rewritten as 
\begin{eqnarray}
    && f \phi''+f' \phi' -z \phi +\partial_{x}^{2} \phi +\left( \frac{A_{t}^{2}}{f} \right)\phi =0, \label{eq:eom1}\\
    && f A_{t}''+\partial_{x}^{2}A_{t}- \phi^{2}A_{t}=0. \label{eq:eom2}
\end{eqnarray}
Though these non-linear partial differential equations are difficult to solve analytically, one can solve them numerically. Note that these equations can give homogeneous solutions of both the normal state and the superconducting state.

First we consider boundary conditions. The gauge field must vanish at the horizon as in the homogeneous case:\,$A_{t}(1,x)=0$. The boundary condition of the scalar field is obtained by imposing the regularity condition to equations of motion at the horizon. 
We consider the case that the chemical potential is constant in space at the AdS boundary:\,$A_{t}(0,x)=\mu$. For the scalar field, we assume the same condition $\Psi^{(1)}=0$ as in the homogeneous case. In terms of $\phi(z,x)$, this condition is corresponding to $\phi(0,x)=0$.

In addition, we need to impose the boundary conditions for the $x$ coordinate. First, we employ the Neumann boundary condition at $x=\pm L/2$:
\begin{equation}
    \partial_{x} \phi(z, x=\pm L/2) =0, \;\;\; \partial_{x} A_{t}(z,x=\pm L/2)=0.
\end{equation}
for technical reason. Then we will see the behavior of the solution with taking $L$ larger. If the solution asymptotes to a fixed profile and is stable against the variation of $L$ at a sufficiently large $L$, we adopt the solution.
In summary, our boundary conditions are following
\begin{eqnarray}
    && (\left. f' \phi'-z \phi +\partial_{x}^{2} \phi \right)|_{z=1}=0, \;\;\; A_{t}(1,x)=0, \\
    && \phi(0,x) =0, \;\;\; A_{t}(0,x)=\mu, \\ 
    && \partial_{x} \phi(z, \pm L/2) =0, \;\;\; \partial_{x} A_{t}(z,\pm L/2)=0.
\end{eqnarray}

In order to perform numerical calculations, we employ the Newton-Raphson relaxation method. In this scheme, we first prepare a seed configuration with $L=4$ and iterate this until it relaxes to a stable solution within a small numerical error.
After we obtain the solution, we examine whether it survives when $L$ becomes sufficiently large. In our study, we confirm this up to $L=40$.
We also introduce a cutoff $\varepsilon=10^{-10}$ at the AdS boundary in order to avoid the numerical divergence. 

We have to prepare an appropriate seed configuration in order to find inhomogeneous solutions. One candidate is an odd function of the spatial coordinate $x$ for the scalar field as discussed in\,\cite{Keranen2009,Keranen22009}, which has a single node on the $x$-axis. After we iterate the relaxation process, we obtain inhomogeneous solutions of $\phi(z,x)$ and $A_{t}(z,x)$ as shown in Fig.\,\ref{fig:3D}. In addition, if we control the number of nodes of an initial seed along the $x$-axis and the value of the chemical potential, we also find other solutions as shown in Figs.\,\ref{fig:3D2} and \ref{fig:3D3}.

\begin{figure}[tbp]
\includegraphics[width=7.5cm,bb=0 0 360 232]{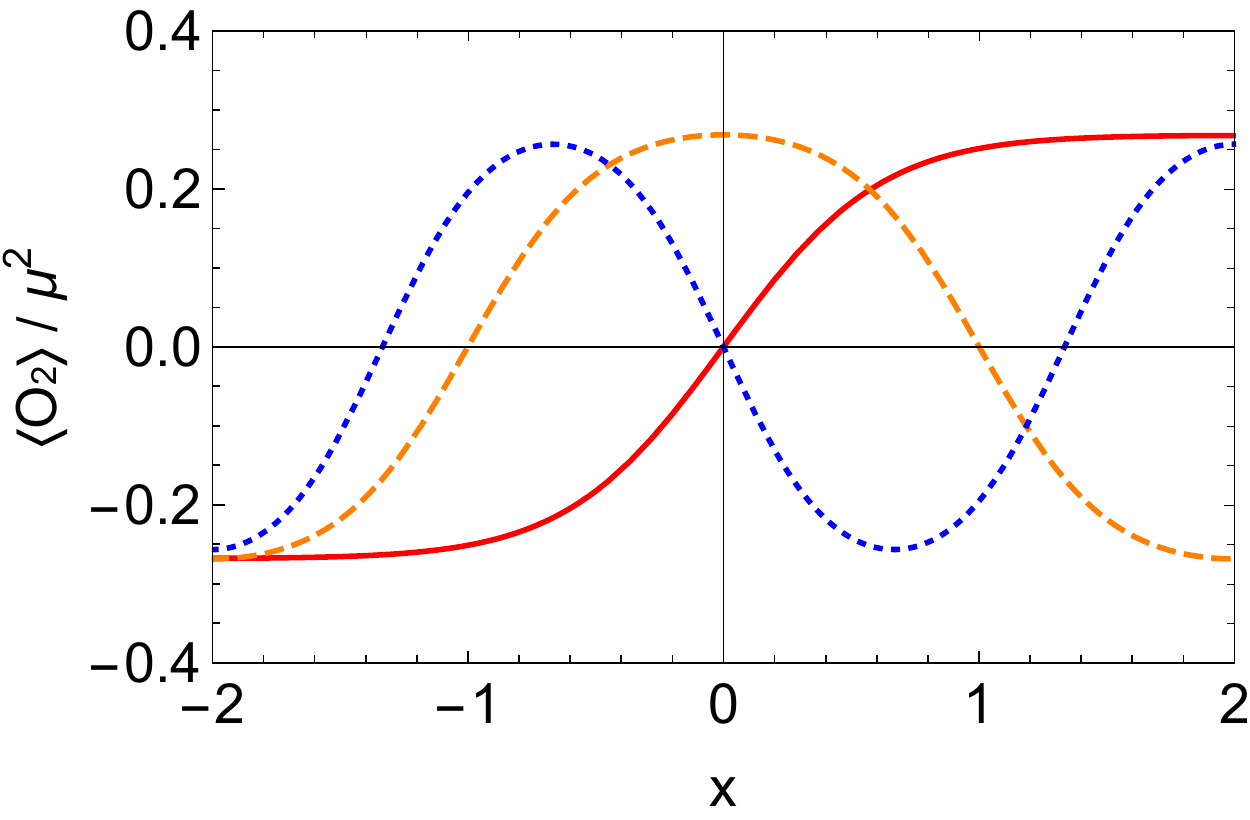}
\includegraphics[width=7.5cm,bb=0 0 360 232]{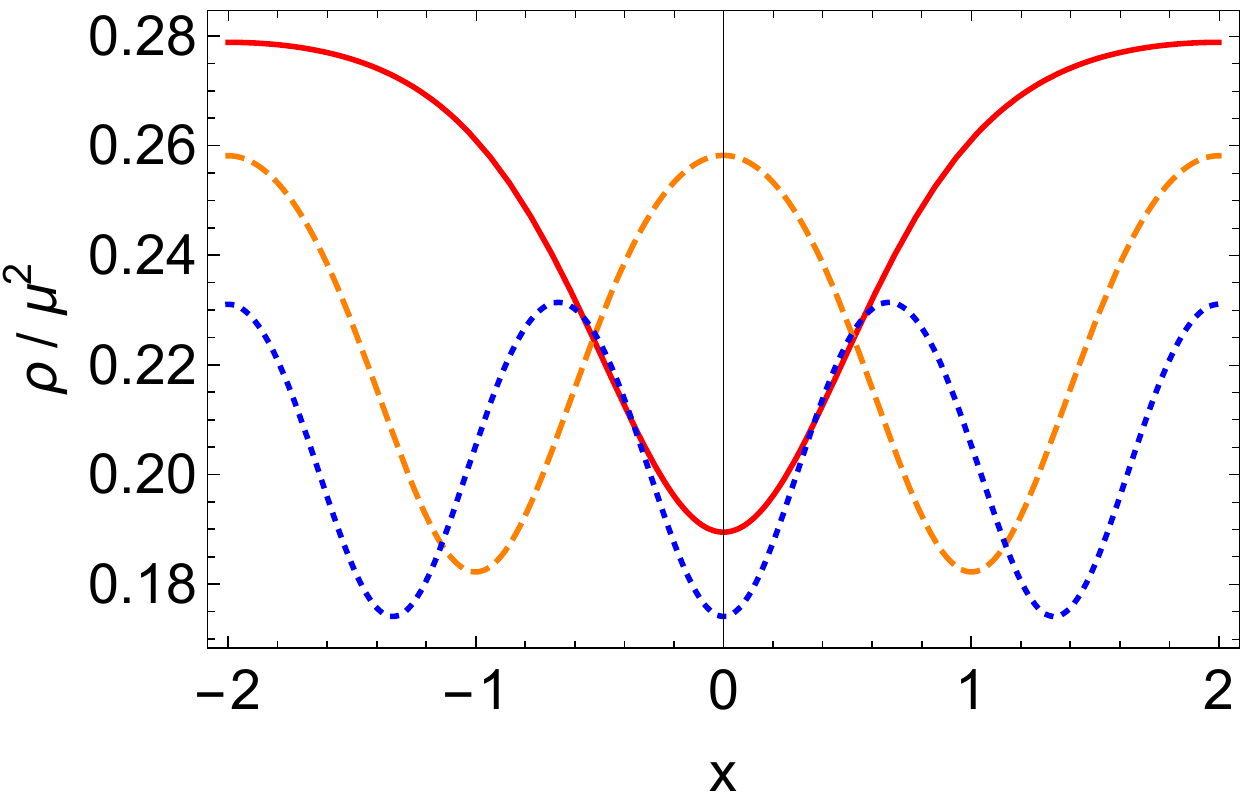}
\caption{The condensate\,(left) and the charge density\,(right) profiles obtained from each bulk solution for $\mu=6.5$\,(solid line), $\mu=7.5$\,(dashed line), and $\mu=8.5$\,(dotted line), respectively.}
\label{fig:op}
\end{figure}

We obtain the condensate $\expval{{\cal O}_{2}}$ and the charge density $\rho$ from the asymptotic form of each field in the vicinity of the AdS boundary,
\begin{equation}
    \phi(z,x)=\expval{{\cal O}_{2}} z + \cdots , \;\;\; A_{t}(z,x) =\mu-\rho z + \cdots.
\end{equation}
Here, it is convenient to normalize the condensate and the charge density by using the chemical potential. In Fig.\;\ref{fig:op}, the condensate $\expval{{\cal O}_{2}}/\mu^{2}$ and the charge density $\rho/\mu^{2}$ profiles obtained from above bulk solutions are shown. We find that solutions with higher number of nodes appear for larger chemical potentials. We also find that dips of the charge density are located at the nodes of the condensate. 

\begin{figure}[tbp]
\includegraphics[width=7.5cm,bb=0 0 360 232]{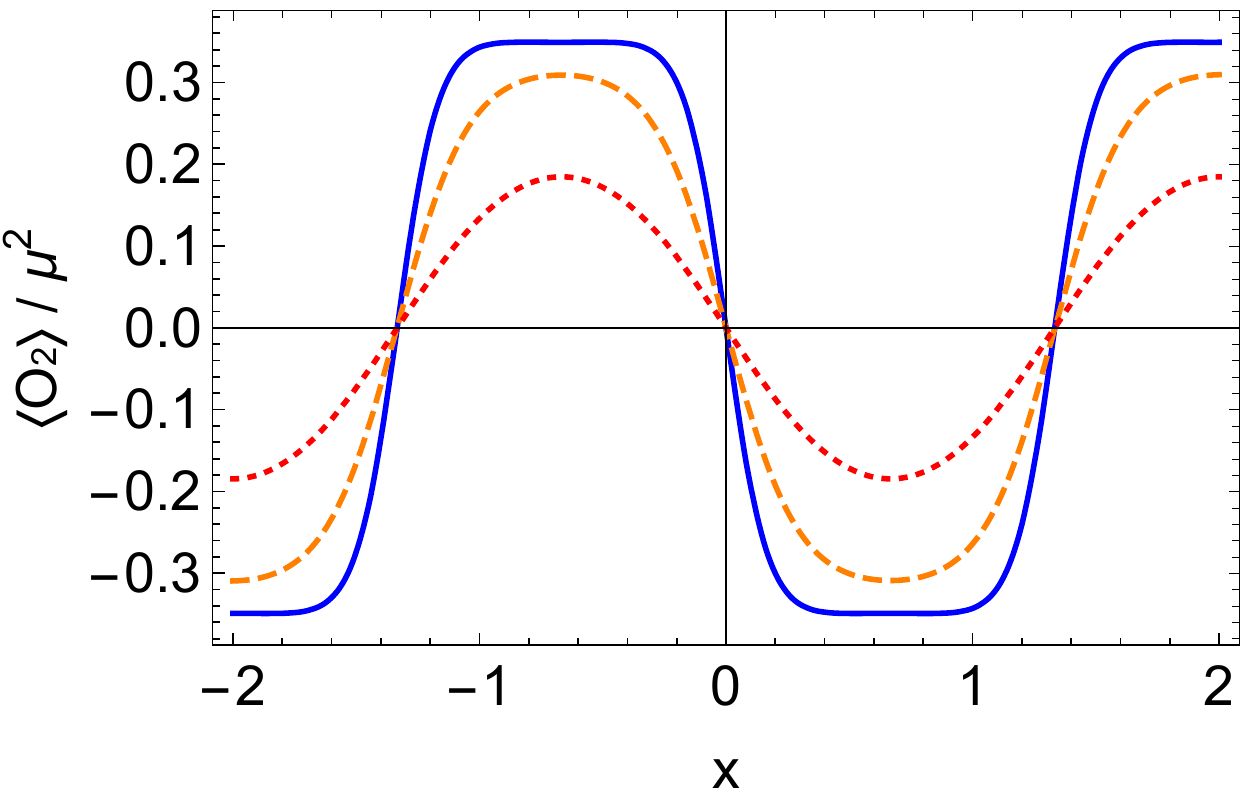}
\includegraphics[width=7.5cm,bb=0 0 360 232]{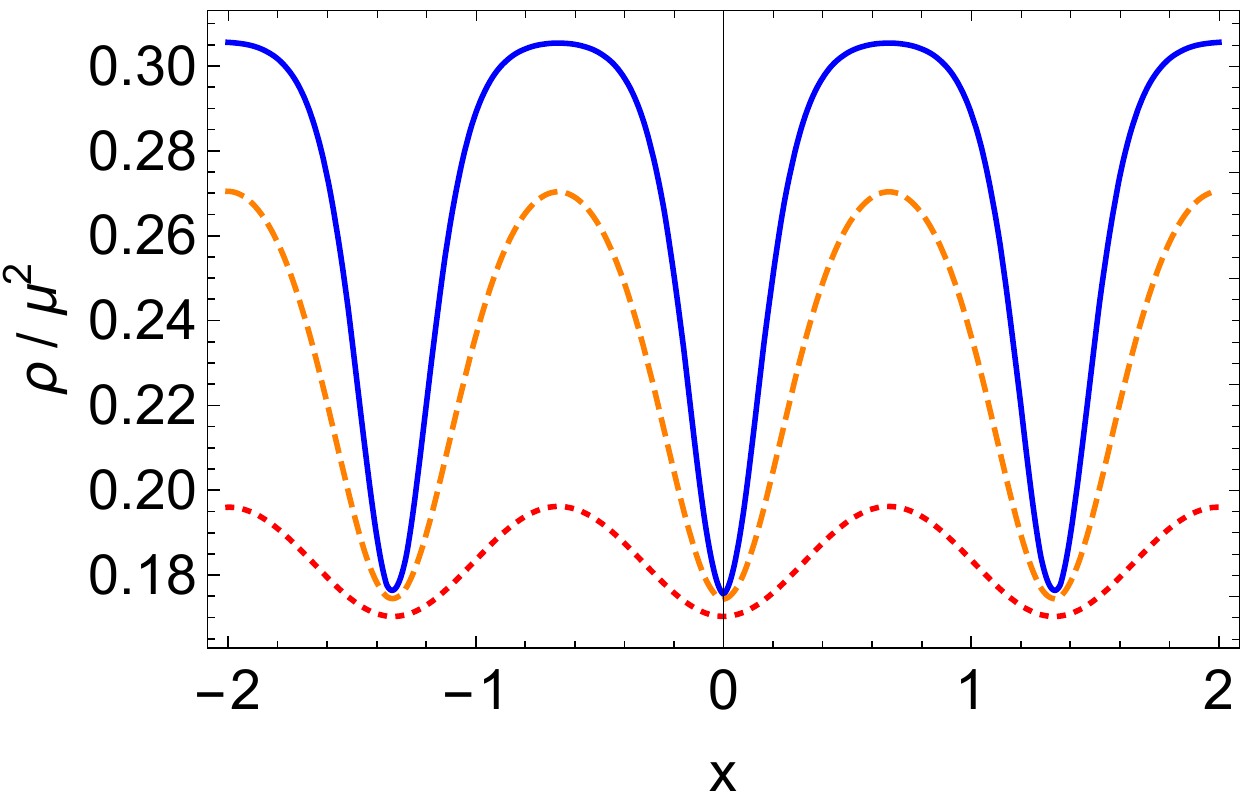}
\caption{The condensate\,(left) and the charge density\,(right) profiles obtained from each bulk solution with three nodes for $\mu=20$\,(solid line), $\mu=12$\,(dashed line), and $\mu=7$\,(dotted line), respectively.}
\label{fig:plotmu}
\end{figure}

\begin{figure}[tbp]
\centering
\includegraphics[width=10cm,bb=20 0 360 237]{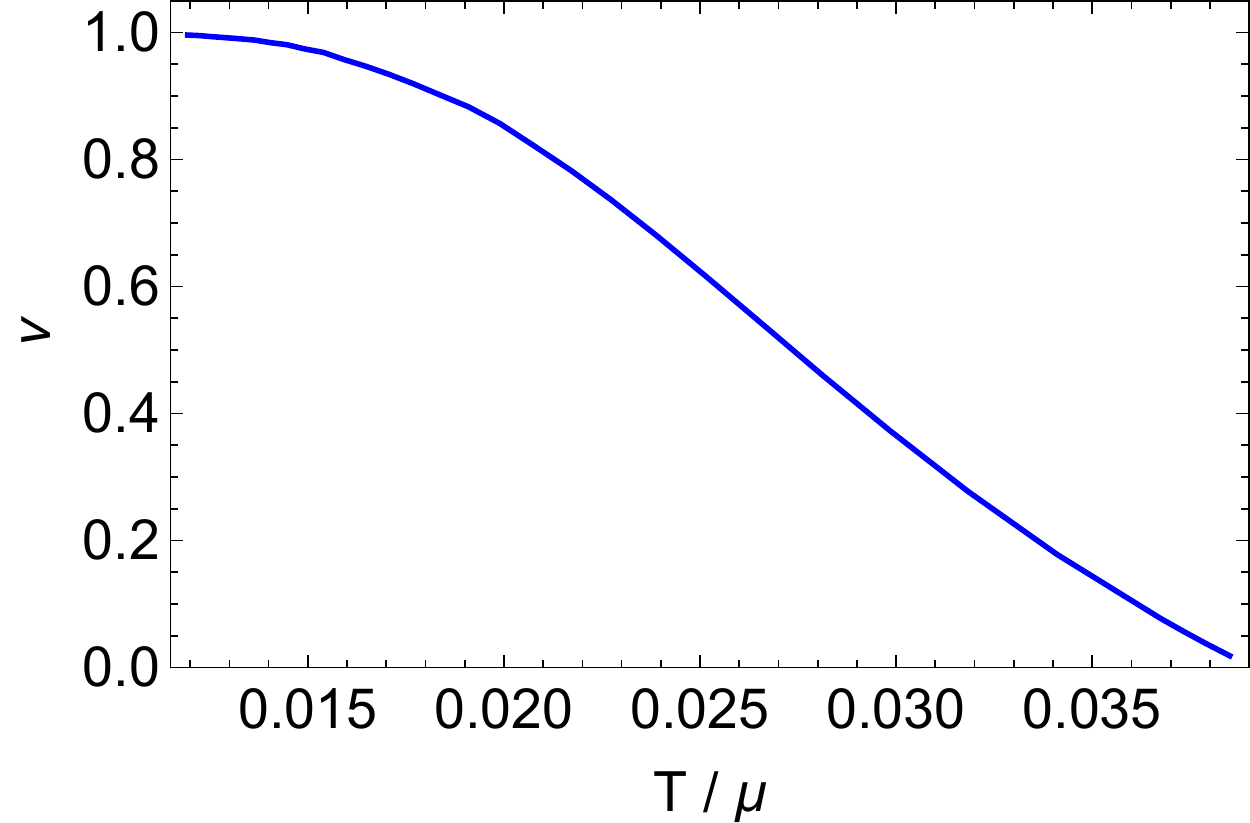}
\caption{The $T/\mu$ dependence of the parameter $\nu$, which characterizes the profile of the Jacobi elliptic functions.}
\label{fig:nu}
\end{figure}

Furthermore, we study the $T/\mu$ dependence of inhomogeneous solutions with the number of nodes fixed. Fig.\,\ref{fig:plotmu} shows the condensate and the charge density profiles with three nodes for each $T/\mu$. These numerical results imply that our inhomogeneous condensates can be characterized by the Jacobi elliptic functions, which is defined by $M(x)=a\,{\rm sn}(b x, \nu)$. Here, $a$ and $b$ are constants and $\nu$ is the so-called elliptic modulus, which determines the profile of functions\,($0\leq \nu \leq 1$).
It has been known that the Jacobi elliptic functions is an analytic solution of the (1+1) dimensional Gross-Neveu model, which exhibits the solitonic inhomogeneous behavior\,(see Refs.\,\cite{Basar2008,Basar2008k} and references therein). Assuming that the condensate profile follows the Jacobi elliptic functions, we determine the $T/\mu$ dependence of $\nu$ by fitting as shown in Fig.\,\ref{fig:nu}.

\begin{figure}[tbp]
\includegraphics[width=7.5cm,bb=0 0 360 232]{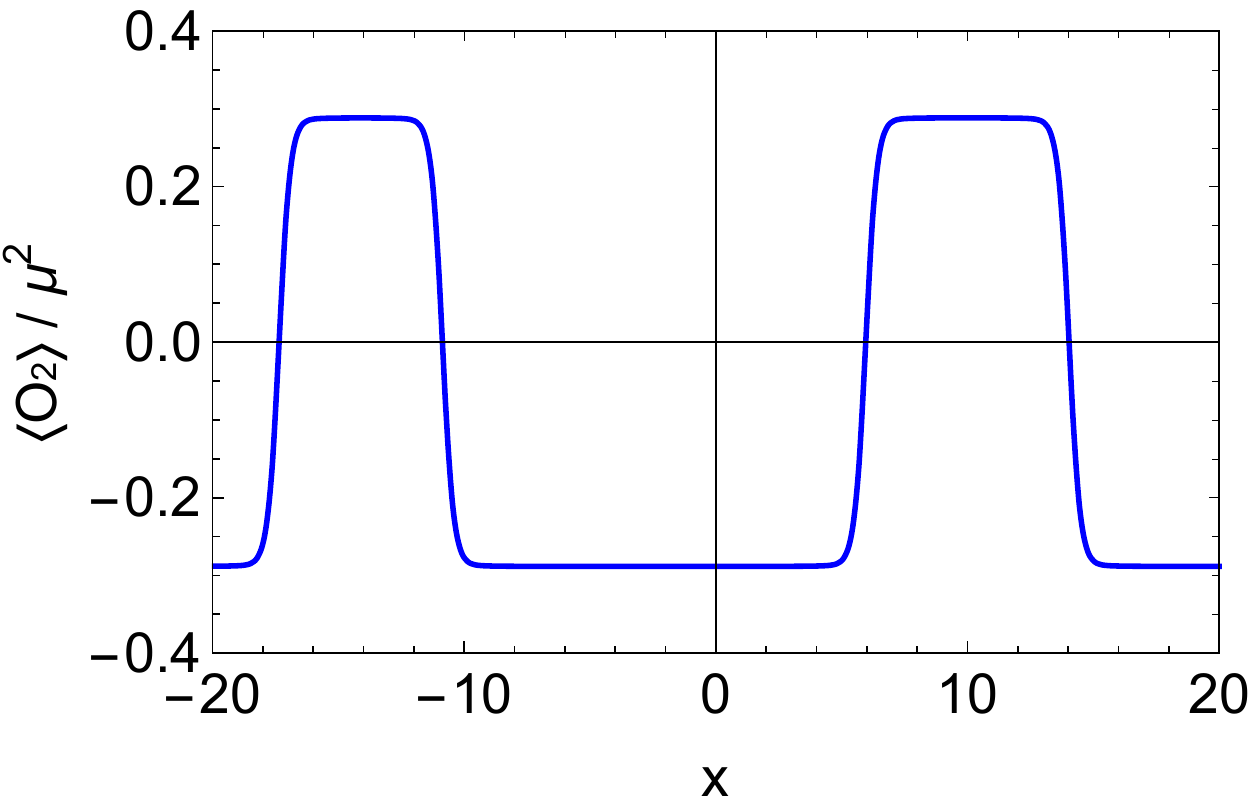}
\includegraphics[width=7.5cm,bb=0 0 360 232]{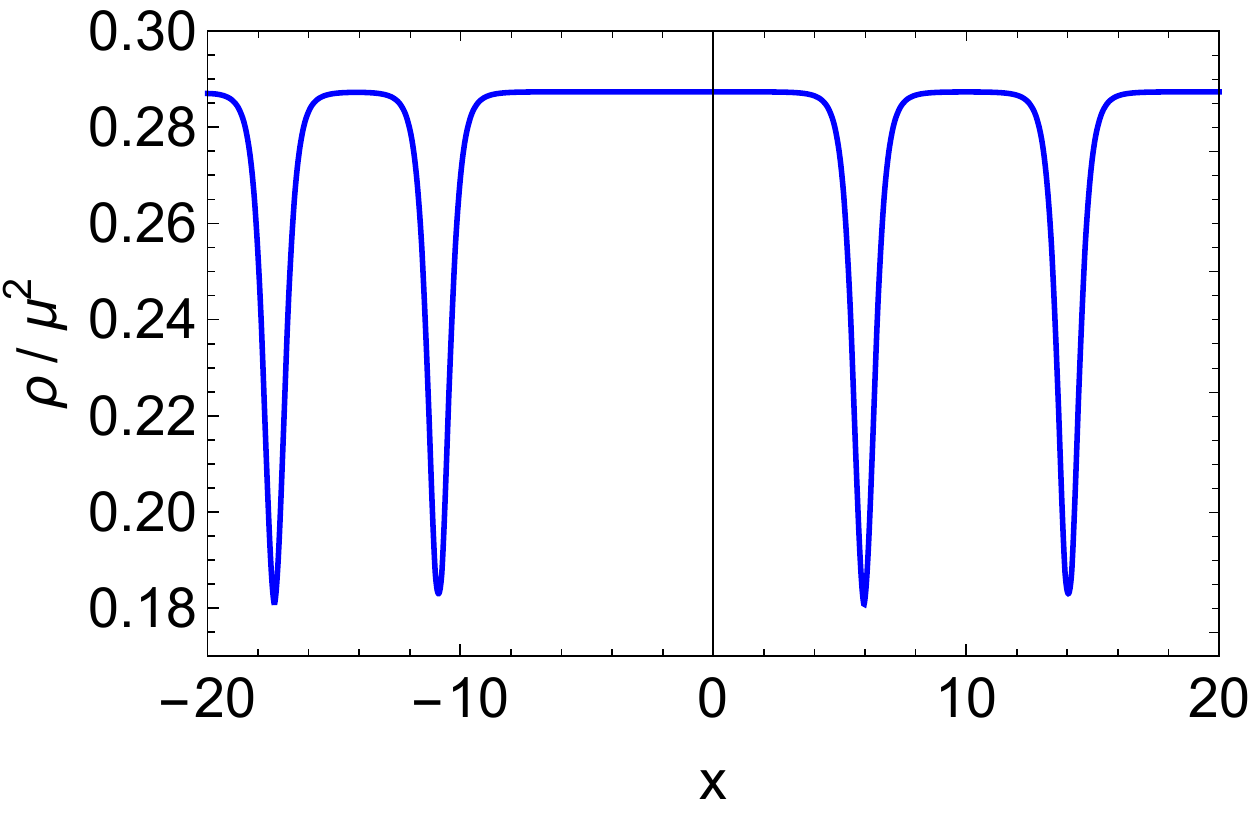}
\caption{The condensate\,(left) and the charge density\,(right) profiles obtained from each bulk solution for $\mu=8$ and $L=40$. Each kink and anti-kink independently exist\,(a dilute gas of kinks and anti-kinks).}
\label{fig:dilute}
\end{figure}

\begin{figure}[tbp]
\includegraphics[width=7.5cm,bb=0 0 360 232]{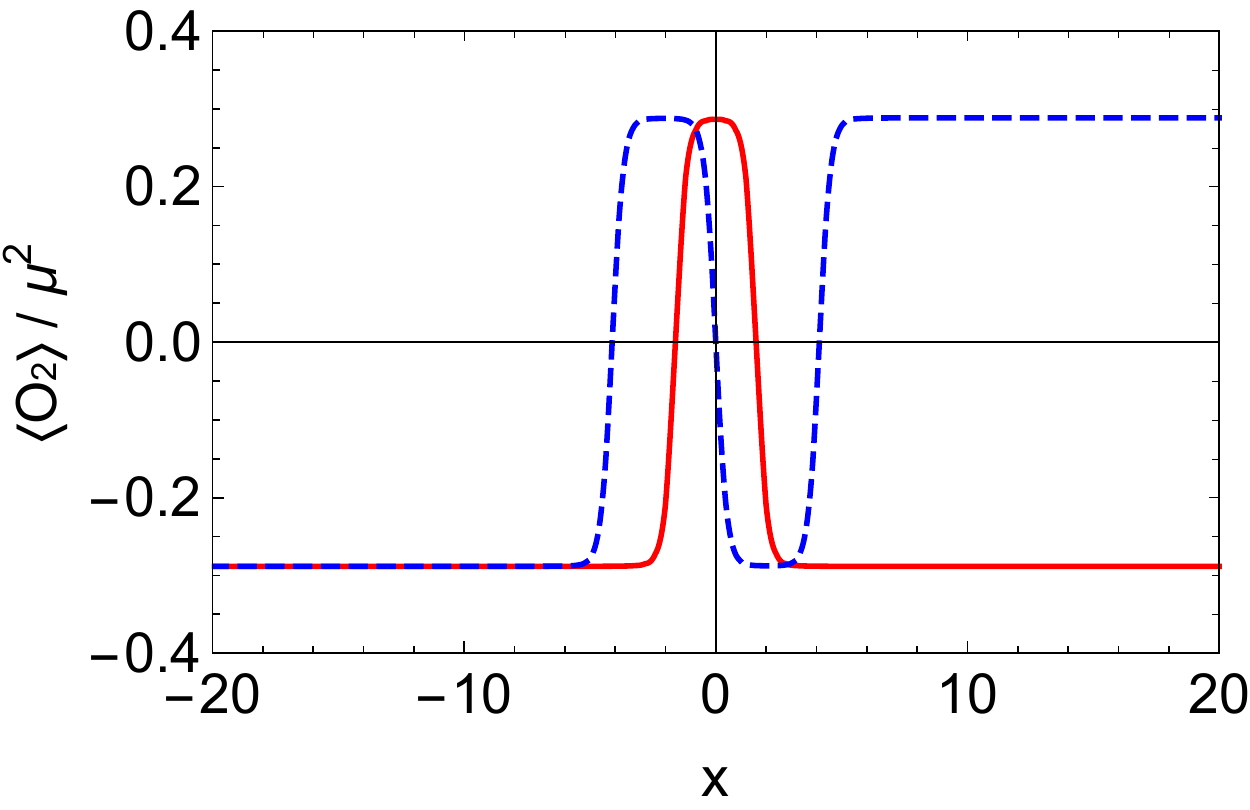}
\includegraphics[width=7.5cm,bb=0 0 360 232]{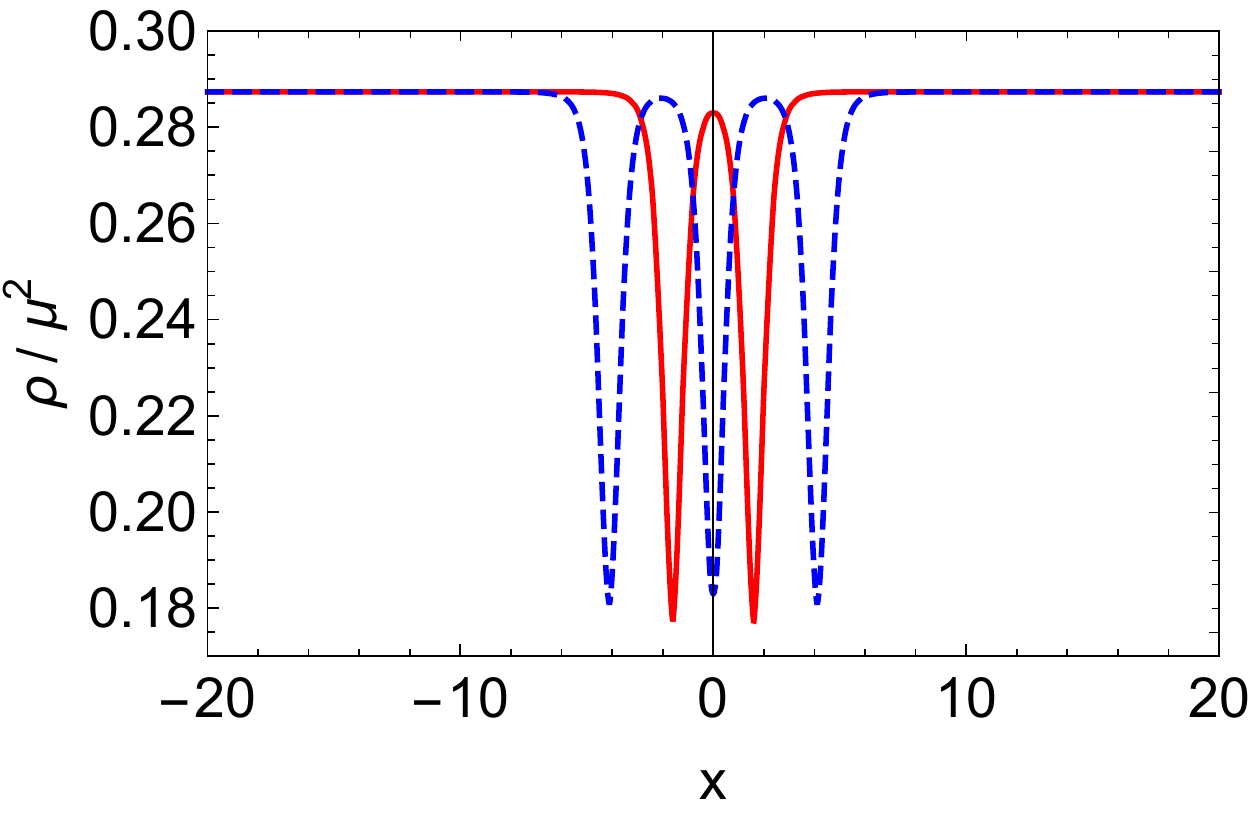}
\caption{The condensate\,(left) and the charge density\,(right) profiles obtained from each bulk solution with two\,(dashed line) or three nodes\,(solid line) for $\mu=8$ and $L=40$. These solutions correspond to the multi-kink solution.}
\label{fig:kink}
\end{figure}

Here, we confirm whether the inhomogeneous solutions survive when $L$ becomes large. We find two categories of solutions. The first category of solutions has a property that the number of nodes increases when $L$ becomes large. These solutions have a periodic structure as already shown in Fig.\,\ref{fig:plotmu}. Thus, we regard this category of solutions as the kink crystalline condensate (LO-like phase). The second category of solutions shows a property that the number of nodes is unchanged when $L$ becomes large. This category of solutions can be further classified into two types. The first type of solutions has a feature that the distance between the nodes can be larger when we increase $L$. Each kink and anti-kink is considered to be independent at sufficiently large $L$, and we regard these solutions as a dilute gas of kinks and anti-kinks. An example of this type of solutions for $L=40$ is shown in Fig.\,\ref{fig:dilute}. The other type of solutions has a notable property that the distance between the nodes is almost unchanged when $L$ becomes large.
We show the condensate and the charge density profiles with two or three nodes for $L=40$ in Fig.\,\ref{fig:kink}. This type of solution corresponds to the multi-kink solution. For example, the condensate with two nodes corresponds to the kink--anti-kink solution.
Surprisingly, we find that the holographic superconductor model achieves to yield the multi-kink solution, which is not the solution of the conventional nonlinear Schr{\"o}dinger equation\,(conventional Ginzburg-Landau theory)\,\cite{Takahashi2012,Correa2009}. It is known that the multi-kink solution is obtained from the Gross-Neveu model or the Ginzburg-Landau theory with higher derivatives and with the higher order potential terms. Thus, the holographic superconductor model produces phenomena beyond the conventional Ginzburg-Landau model.

\section{Thermodynamic Potential}
\label{sec:3}

\begin{figure}[tbp]
\centering
\includegraphics[width=12cm,bb=0 0 360 297]{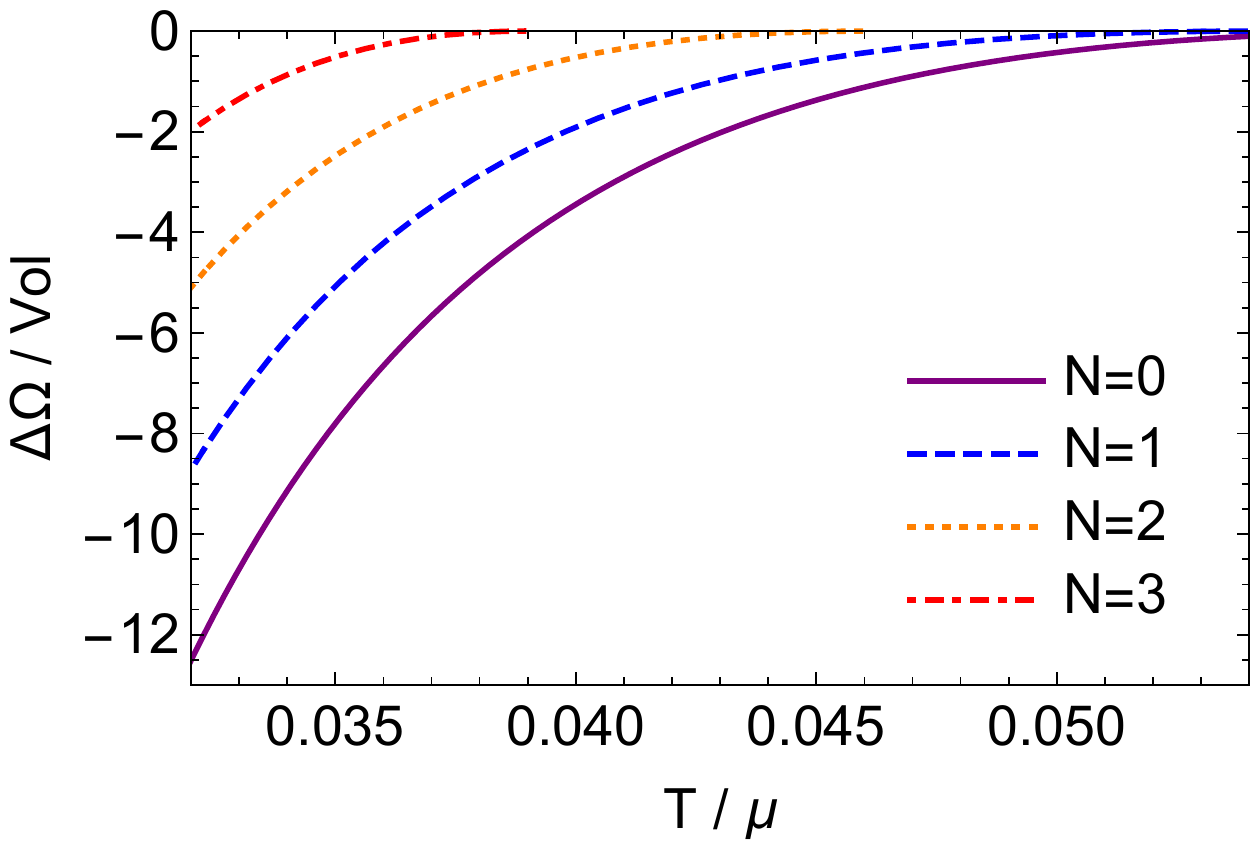}
\caption{The thermodynamic potential that has each number of nodes. $N$ is the number of nodes.}
\label{fig:TP}
\end{figure}

In this section, we study the thermodynamic potential for inhomogeneous superconducting states in order to analyze their stability.
According to the AdS/CFT dictionary, the thermodynamic potential of the field theory on the boundary corresponds to $-L_{\rm os}$, where $L_{\rm os}$ is the on-shell Lagrangian. We derive the thermodynamic potential for our inhomogeneous superconducting state following Ref.\,\cite{Flauger2010}. We rewrite the action (\ref{eq:action}) as the explicit form
\begin{equation}
    S=\int d^{4}x \left\{ \frac{1}{2} \left[ (A_{t}')^{2}+\frac{(\partial_{x} A_{t})^{2}}{f} \right] -\frac{1}{z^{2}}\left[ f(\Psi')^{2}+(\partial_{x}\Psi)^{2}\right]+\frac{\Psi^{2}}{z^{4}}\left(\frac{A_{t}^{2}z^{2}}{f}+2 \right) \right\}.
    \label{eq:action2}
\end{equation}
The on-shell action is obtained by integrating by parts and substituting equations of motion (\ref{eq:eom1}) and (\ref{eq:eom2}) into Eq.\,(\ref{eq:action2}).
\begin{equation}
    S_{\rm os}=\left. \int d^{3}x \left(\frac{f}{z^{2}}\Psi \Psi' -\frac{1}{2}A_{t}A_{t}' \right) \right|_{z=0}-\int d^{4}x \frac{A_{t}^{2}}{z^{2}f}\Psi^{2},
    \label{eq:Sonshell}
\end{equation}
where we have also used boundary conditions given in the previous section and $f(1)=0$. In this study, furthermore, since we assume that $\Psi^{(1)}=0$ in (\ref{eq:asymptotic}), $\Psi \Psi' /z^{2}$ in Eq.\,(\ref{eq:Sonshell}) vanishes. Hence, the thermodynamic potential is obtained as
\begin{equation}
    \Omega=\left. \int d^{2}x \frac{1}{2}A_{t}A_{t}' \right|_{z=0} +\int d^{3}x \frac{A_{t}^{2}}{2 f}\phi^{2},
\end{equation}
where we change the notation of the scalar field:\,$\Psi\rightarrow z \phi /\sqrt{2}$. The thermodynamic potential per unit volume is given by
\begin{equation}
    \frac{\Omega}{\rm Vol}=\frac{1}{L}\left( \left. \int_{-L/2}^{L/2} dx \frac{1}{2}A_{t}A_{t}' \right|_{z=0} +\int dx dz \frac{A_{t}^{2}}{2 f} \phi^{2} \right),
\end{equation}
where ${\rm Vol}=L \int dy$. 
In Fig.\,\ref{fig:TP}, we show $\Delta \Omega / {\rm Vol}$ as a function of $T/\mu$ for kink crystalline solutions whose number of nodes is denoted by $N$, where $\Delta \Omega$ is the difference of the thermodynamic potential between the normal state and the superconducting state. We find that kink crystalline solutions that have higher number of nodes are realized in lower $T/\mu$, namely lower $T$ and/or larger $\mu$. Moreover, we find the inhomogeneous solutions have lower $\Delta \Omega$ than the normal state, but higher than homogeneous solutions. In other words, the inhomogeneous solutions that we found in this study are metastable. We confirm that this qualitative behavior remains unchanged when $L$ becomes sufficiently large.
\section{Conclusion and Discussion}
\label{sec:5}
In this paper, we have explored the inhomogeneous solutions of the holographic superconductor to investigate what types of solutions are contained in the model.
We find that the three types of solutions of the condensate:\,the kink crystalline condensate (LO-like phase), the dilute gas of kinks and anti-kinks, and the multi-kink solution.
It is known that the LO phase and the dilute gas of kinks and anti-kinks are realized in the conventional Ginzburg-Landau model with $\phi^{4}$ potential. On the other hand, this conventional Ginzburg-Landau model does not have the multi-kink solution, for instance, the kink--anti-kink solution. Evaluating the thermodynamic potential, we find that these inhomogeneous solutions are metastable. It is known that the multi-kink solutions of the microscopic theories such as the BCS model, or the Gross-Neveu model are also metastable. Therefore, our results show that the holographic superconductor model exhibits the physics of superconductor, which appears in the microscopic models, beyond the conventional Ginzburg-Landau model.

We confirm that the condensate with higher number of nodes emerges at lower $T/\mu$. In the LO phase, the profile of the condensate can be characterized by the Jacobi elliptic functions. A similar discussion is given in QCD\,\cite{Casalbuoni2003}. Assuming that the condensate profile with multiple nodes follows the Jacobi elliptic functions, we determine the $T/\mu$ dependence of $\nu$ by fitting. We reveal that the value of $\nu$ approaches asymptotically to $\nu=1$ at smaller $T/\mu$. 

Before closing the conclusion, we have some remarks. First, we are interested in the connection between the holographic kink crystalline condensate found in this work and the LOFF phase. 
In the LOFF phase, the spin degree of freedom plays central role for the stabilization of the LOFF phase. Therefore, it would be interesting to apply an external magnetic field so that we can access the nature of the possible spin degree of freedom in the model.
Moreover, it is also known that Aharonov-Bohm flux increases the critical temperature of the LOFF phase\,\cite{Zyuzin2008,Zyuzin2009}. It will be straightforward to apply our method to these setups.
These perspectives are related to another interesting question of whether the inhomogeneous solution in holography can become the ground state\footnote{It has been reported that inhomogeneous solutions are realized in various setups, for instance, as shown in Refs.\,\cite{Cremonini2016,Nakamura2009,Donos2011}}. It can be expected that inhomogeneous solutions are energetically favored with a different form of the potential, for example, which includes higher derivative terms as found in\,\cite{Nickel2009}.
Since we have restricted our analysis to the probe limit, we ignore the backreaction of the matter sector into the gravity sector, which becomes important at low temperature. It is straightforward to consider the backreaction in our case by following recent work\,\cite{Xu2019}. 
This study partially reveals that the holographic superconductor model has a richer structure than the conventional Ginzburg-Landau theory. It would be interesting to investigate how the highly non-trivial inhomogeneous solutions, such as the multi-kink solution realize instead of the simple setup in the holographic superconductor model and that would lead us to reveal a cornucopia of new physics.

\section*{Acknowledgement}
The authors are grateful to S.\,Kinoshita for helpful discussions and comments. The work of M.\,M. is supported by the Research Assistant Fellowship of Chuo University. 
The work of S.\,N. is supported in part by JSPS KAKENHI Grant Numbers JP19K03659, JP19H05821, and the Chuo University Personal Research Grant.
The work of R.\,Y. is supported in part by JSPS KAKENHI Grant Number JP19K14616.


\end{document}